\documentclass[epsfig]{article}
\usepackage{glas,epsfig}
\newlength{\dinwidth}
\newlength{\dinmargin}
\newcommand{\AmS}{{\protect\the\textfont2
  A\kern-.1667em\lower.5ex\hbox{M}\kern-.125emS}}

\def\pl#1#2#3   {{\em Phys. Lett.} {\bf#1} (#2) #3}
\def\eurj#1#2#3 {{\em Eur. Phys. J.} {\bf#1} (#2) #3}
\def\np#1#2#3   {{\em Nucl. Phys.} {\bf#1} (#2) #3}
\def\NIM#1#2#3 {{\em Nucl. Instrum. Methods} {\bf#1} (#2) #3}
\def\prev#1#2#3 {{\em Phys. Rev.} {\bf#1} (#2) #3}
\def\cpc#1#2#3  {{\em Computer Phys. Comm.} {\bf#1} (#2) #3}

\begin{document}
\begin{titlepage}{GLAS-PPE/1999--06}{May 1999}

\title{Structure of Real and Virtual Photons from ZEUS}

\author{N. Macdonald}


\begin{abstract}
Measurements sensitive to the structure of both real and virtual photons are
presented and compared to theoretical models
with various photon parton distribution functions (PDFs). Measurements for
real photons show a tendency for the available photon PDFs to be too
small to describe the data. For virtual photons, the photon PDF is
seen to decrease with increasing photon virtuality. In order to
describe the data, resolved photon processes are required up to a
photon virtuality of at least 4.5 GeV$^2$.

\end{abstract}

\end{titlepage}


\section{Introduction}

Experimental information on the partonic structure of the photon can
be obtained from the data taken at the HERA ep
collider experiments. 
Leading order (LO) QCD predicts that photon interactions
have a two-component nature.
In direct photon processes the entire momentum of the photon takes part in
the hard subprocess with a parton from the proton
whereas in resolved photon processes the photon acts as a source of
partons and one of these enters the hard subprocess.
By measuring inclusive dijet events (two or more jets) information on
the structure of the real photon can be extracted. By also measuring the
scattered electron, information can be obtained on the evolution of 
this structure as a function of the virtuality of the photon, Q$^2$.

The fraction of the photon's four momentum which enters the hard 
subprocess at leading order, denoted by $x_\gamma^{LO}$, is equal to unity
for direct processes, and less than unity for resolved processes. 
Experimentally it is not possible to measure
$x_\gamma^{LO}$ directly. 
Instead, an observable quantity $x_\gamma^{obs}$ is defined which is
calculable and well-defined to all orders of perturbation theory. 
$x_\gamma^{obs}$ is the fraction of the photon momentum manifest in the two
highest transverse energy jets and is defined by the equation
\begin{eqnarray}
x_\gamma^{obs} = \frac{\sum E_{Tj} e^{- \eta_j}}{2E_e y} \nonumber
\end{eqnarray}
where $E_{Tj}$ is the transverse energy of jet $j$, $\eta_j$ is the
pseudorapidity of the jet measured in the lab frame, 
and $y$ is the inelasticity of the event.
\par
``Direct enriched'' events are defined as being those with
$x_\gamma^{obs} > 0.75$ and ``resolved enriched'' events as those with
$x_\gamma^{obs} < 0.75$. This value gives the optimal separation of
the leading order direct and resolved event classes.

\section{Dijets in Photoproduction and Real Photon Structure}

The kinematic selection cuts made in order to examine the structure of
the real photon are
\begin{itemize}
        \item Two or more jets ($k_T$ clustering algorithm)
        \item Q$^{2} \simeq 0$ GeV$^2$
        \item $E_{T~leading}^{jet} > 14$ GeV, $E_{T~second}^{jet} > 11$ GeV
        \item $-1 < {\eta}^{jet} < 2$
        \item $0.20 < y < 0.85$
\end{itemize}
\par
The advantages of using high $E_{T}$ dijets are that they provide a
hard scale where perturbative QCD (pQCD) is expected to work, the
hadronisation corrections are small, and the effect of underlying
events is small. Given these assumptions, the data can be compared
directly to NLO pQCD calculations without the need to simulate some
hadronisation model. This analysis 
concentrates on the high $x_{\gamma}^{obs}$ region where the quark 
densities in the photon are not strongly constrained by 
$e^{+}e^{-}$ experiments. 

The measured ZEUS data is compared to NLO pQCD calculations for
three photon PDFs,  GRV-HO \cite{grvho}, AFG-HO \cite{afgho} and
GS96-HO \cite{gs96ho}. The NLO calculations have been performed by
three groups of theorists, Harris et al. \cite{harris}, Klasen et
al. \cite{klasen} and Frixione et al. \cite{frixione}. Since the
agreement between the calculations is excellent only one of the
calculations is plotted. 

\begin{figure}[htb]
\centering
\epsfig{file=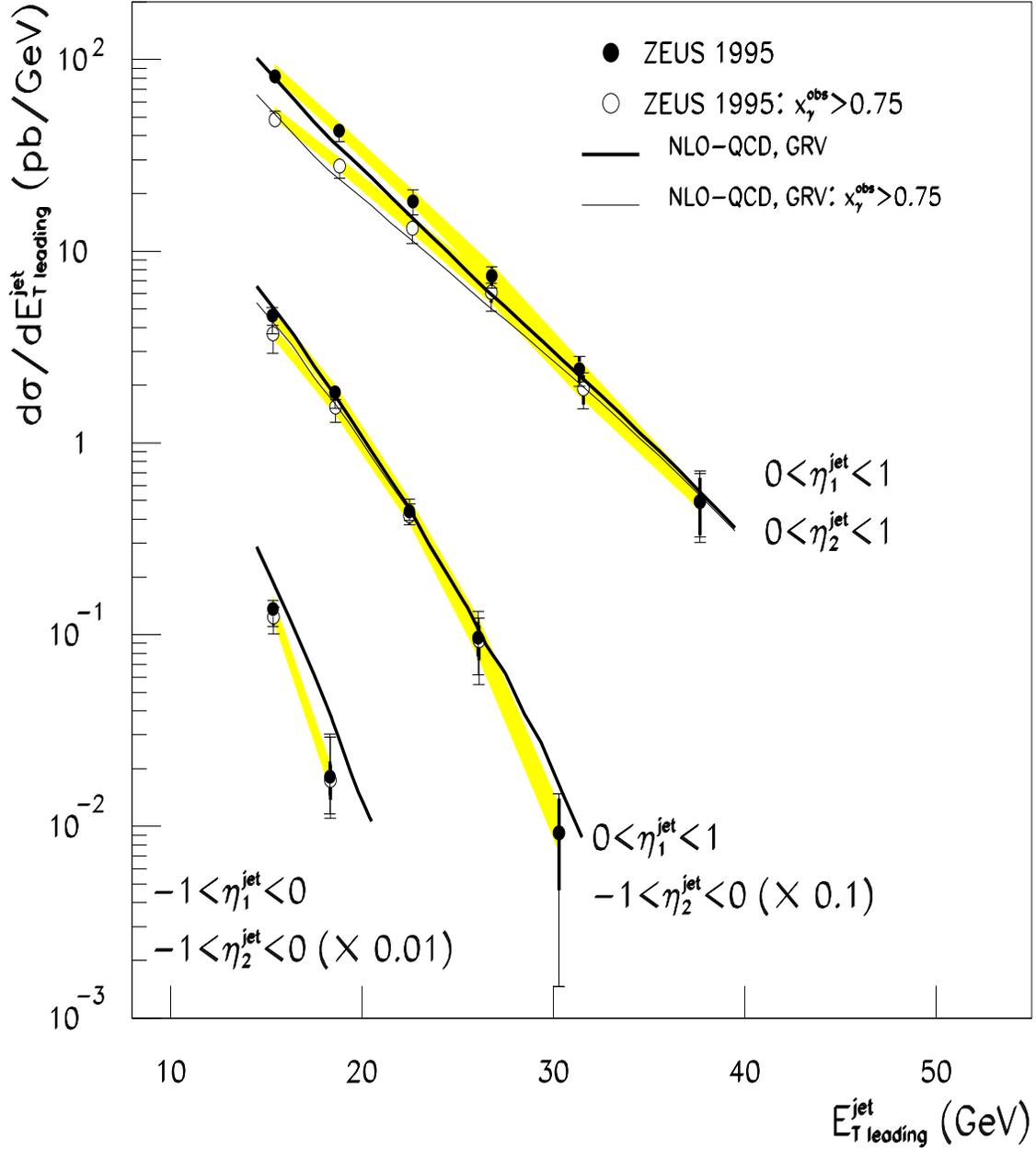,width=15.0cm,height=18.0cm,angle=0}
\caption{Differential dijet cross section as a function of
$E_{T~leading}^{jet}$ for 0.20 $< y <$ 0.85.}
\label{et_fully_both_4b}
\end{figure}

Figure~\ref{et_fully_both_4b} shows the differential dijet cross
section as a function of $E_{T~leading}^{jet}$ for 0.20 $< y <$
0.85. There is an excess in the data above
theory for central jets (0 $< \eta^{jet} <$ 1) below an $E_T$ of
25~GeV. The assumption that 
the hadronisation corrections are negligible is not true for
backward jets (-1 $< \eta^{jet} <$ 0), so no conclusions are drawn
about the backward 
region, rather the discrepancy between data and theory is ascribed to
a theoretical uncertainty. 

\begin{figure}[htb]
\centering
\epsfig{file=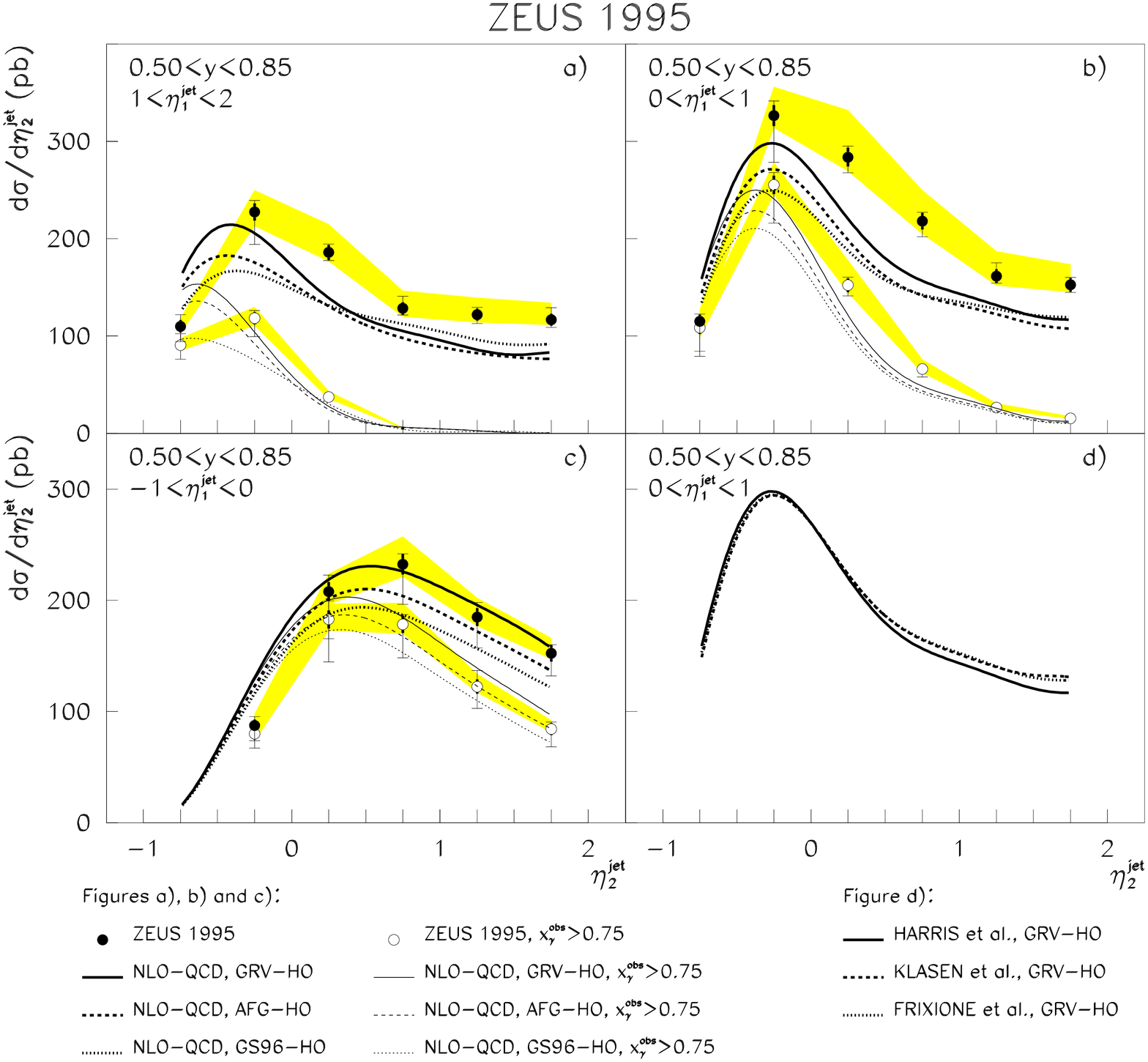,width=15.0cm,height=18.0cm,angle=0}
\caption{Differential dijet cross section as a function of
${\eta}_{2}^{jet}$ in bins of ${\eta}_{1}^{jet}$ for 0.50 $< y <$ 0.85.}
\label{eta_highy_both_32}
\end{figure}

Figure \ref{eta_highy_both_32} shows the cross section for a narrower 
range in $y$. This provides a better sensitivity to the photon
structure since the cross section is no longer averaged over the
broader $y$ range. There is an excess in the data seen for central (0
$< \eta^{jet} <$ 1) and 
forward (1 $< \eta^{jet} <$ 2) jets, both for the whole
$x_\gamma^{obs}$ range and the high 
$x_\gamma^{obs}$ range. 
\par
The excess in the data above theory for jets with $E_{T} < 25$~GeV and
for central and forward rapidities suggests that the available photon
PDF parametrisations, based on $e^{+}e^{-}$ $F_{2}^{\gamma}$ data, are
too small in this kinematic regime. 

\section{Ratio of Dijet Cross Sections vs Q$^2$}

The kinematic selection cuts made for the analysis of virtual photons are
\begin{itemize}
    \item Two or more jets ($k_T$ clustering algorithm)
    \item Q$^{2} \simeq 0$, $0.1 < Q^{2} < 0.55$, $1.5 < Q^{2} < 4.5$ (GeV$^2$)
    \item $0.20 < y < 0.55$
    \item $E_{T}^{jets} > 5.5$ GeV
    \item $-1.125 < {\eta}^{jets} < 2.2$
\end{itemize}
\par
Three different Q$^2$ regions are available for measurement. 
Q$^{2} \simeq 1.0$
GeV$^2$ corresponds to quasi-real photons, where the electron is not
measured, and the large bulk of such events lead to a median Q$^2$ of
0.001 GeV$^2$. ZEUS has a small angle
electron detector for tagging events in the 
transition region between
photoproduction and DIS ($0.1 < Q^{2} < 0.55$ GeV$^2$). For $1.5 <
Q^{2} < 4.5$ GeV$^2$ the electron 
is detected in the ZEUS main calorimeter.

\begin{figure}[htb]
\centering
\epsfig{file=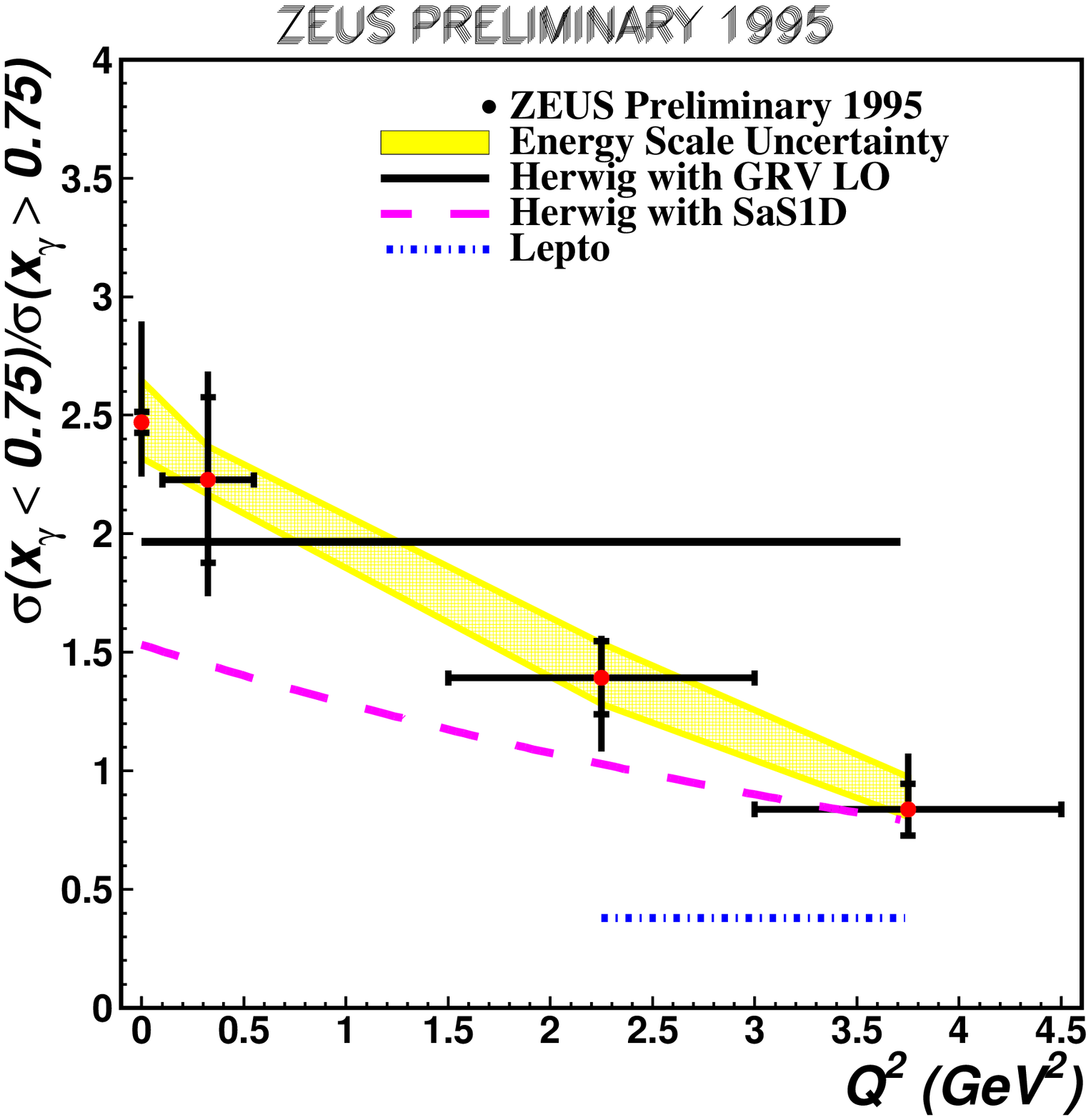,width=15.0cm,height=18.0cm,angle=0}
\caption{Ratio of resolved enriched to direct enriched dijet cross
sections vs Q$^2$}
\label{plotb}
\end{figure}

Figure~\ref{plotb} shows the ratio of the resolved to the direct dijet cross
sections as a function of Q$^2$.  Since some of the systematic
uncertainties cancel in the ratio, this is a more precise
measurement than an absolute differential cross section.  The data 
fall with increasing Q$^2$.  This is compared to the dijet ratio
obtained with HERWIG 5.9 \cite{HERWIG} using a photon PDF which 
does not evolve with Q$^2$ (GRV LO \cite{GRV}) and one which does
(SaS1D \cite{sas}).  The HERWIG ratio 
with GRV LO is flat, while that obtained with SaS 1D falls with
increasing Q$^2$.  Therefore the fall of the ratio with Q$^2$ observed
in the data indicates that the data require a virtual photon PDF
which evolves with photon virtuality.  The LEPTO 6.5.1 \cite{lepto} 
curve shows the prediction for leading order direct processes in DIS
only.  Note that this ratio is non-zero as may be understood given
that LEPTO contains higher-order processes in the approximation of
parton-showering 
and hadronization. The data approach this direct only limit, 
however resolved processes are still required in order to describe 
the data up to at least Q$^2$ = 4.5 GeV$^2$.

\begin{thebibliography}{9}

\bibitem{harris} B.W. Harris and J.F. Owens, \prev{D56}{1997}{4007}.

\bibitem{klasen} M. Klasen and G. Kramer, {\em Z. Phys} {\bf C76}
(1997) 67; \\
                 M. Klasen, T. Kleinwort and G. Kramer, \eurj{C1}{1998}{1}.

\bibitem{frixione} S. Frixione and G. Ridolfi, \np{B507}{1997}{315};
\\
                   S. Frixione, \np{B507}{1997}{295}.


\bibitem{grvho} M. Gl\"uck, E. Reya and A. Vogt,
\prev{D45}{1992}{3986}; \\
                M. Gl\"uck, E. Reya and A. Vogt,
\prev{D46}{1992}{1973}. 

\bibitem{gs96ho} L.E. Gordon and J.K. Storrow, \np{B489}{1997}{405}.

\bibitem{afgho} P. Aurenche, J. Guillet and M. Fontannaz, {\em
Z. Phys} {\bf C64} (1994) 621.

\bibitem{HERWIG} HERWIG 5.9; G. Marchesini et al., \cpc{67}{1992}{465}.

\bibitem{GRV} M. Gluck, E. Reya and A. Vogt, \prev{D46}{1992}{1973}.

\bibitem{lepto} G.Ingelman, A. Edin and J. Rathsman, \cpc{101}{1997}{108-134}.

\bibitem{sas} G.Schuler and T. Sj\"ostrand, \pl{B376}{1996}{193}.



\end{thebibliography}
\end{document}